%% file: Hristov_RECOMB2020.tex
\documentclass[pdftex,twoside,11pt,letterpaper]{article}

\skip\footins=4ex						  
\usepackage{amsmath, amsfonts}
\usepackage[sort&compress]{natbib}
\bibpunct{[}{]}{,}{n}{}{} 
\usepackage[pdftex]{graphicx}
\usepackage{caption}
\usepackage{subcaption}
\usepackage{wrapfig}
\usepackage{floatflt}
\usepackage{float}
\usepackage{calc}
\usepackage{color}
\usepackage{floatflt}
\restylefloat{figure}

\usepackage{times}

\newcommand{\comment}[1]{}

\newcommand{\beginsupplement}{
        \setcounter{table}{0}
        \renewcommand{\thetable}{S\arabic{table}}
        \setcounter{figure}{0}
        \renewcommand{\thefigure}{S\arabic{figure}}
}

\begin{document}


\title{A guided network propagation approach to identify disease genes that combines prior and new information}

\author{Borislav H. Hristov, Bernard Chazelle and Mona Singh\footnote{Department of Computer Science and Lewis-Sigler Institute for Integrative Genomics, Princeton University}~\footnote{Email mona@cs.princeton.edu}}
\date{\vspace{-5ex}}

\maketitle
\thispagestyle{empty}
\input{abstract}
\newpage
\input{intro}

\input{methods}

\input{results}

\input{discussion}

\newpage

\small
\bibliographystyle{myrefstyle}
\bibliography{mybib}
\input{supplement}

\newpage

\end{document}

%% file: abstract.tex

\begin{abstract}

A major challenge in biomedical data science is to identify the causal
genes underlying complex genetic diseases. Despite the massive influx
of genome sequencing data, identifying disease-relevant genes remains
difficult as individuals with the same disease may share very few, if
any, genetic variants.  Protein-protein interaction networks provide a
means to tackle this heterogeneity, as genes causing the same disease
tend to be proximal within networks.  Previously, network propagation
approaches have spread ``signal'' across the network from either known
disease genes \emph{or} genes that are newly putatively implicated in
the disease (e.g., found to be mutated in exome studies or linked via
genome-wide association studies).  Here we introduce a general
framework that considers both sources of data within a network context.
Specifically, we use prior knowledge of disease-associated genes to
guide random walks initiated from genes that are newly identified as
perhaps disease-relevant.  In large-scale testing across 24 cancer
types, we demonstrate that our approach for integrating both prior and
new information not only better identifies cancer driver genes than
using either source of information alone but also readily outperforms
other state-of-the-art network-based approaches.  To demonstrate the
versatility of our approach, we also apply it to genome-wide
association data to identify genes functionally relevant for several complex
diseases.  Overall, our work suggests that guided network propagation
approaches that utilize both prior and new data are a powerful means
to identify disease genes.

\end{abstract}
	




%% file: intro.tex
\clearpage
\pagenumbering{arabic}
\section*{Introduction}

Large-scale efforts such as the 1000 Genomes
Project~\cite{10002015global}, The Cancer Genome Atlas (TCGA)~\cite{tcga},
and the Genome Aggregation Database~\cite{Karczewski531210}, among others, have catalogued millions of variants
occurring in tens of thousands of healthy and disease genomes.
Despite this abundance of genomic data, however, understanding the
genetic basis underlying complex human diseases remains  challenging~\cite{kim2013bridging}.  In contrast to simple
Mendelian diseases, for which a small set of commonly shared genetic
variants are responsible for disease phenotypes, complex heterogeneous
diseases are driven by a myriad of combinations of different
alterations.  Individuals exhibiting the same phenotypic outcome---a
particular disease---may share very few, if any, genetic variants,
thereby making it difficult to discover which of numerous 
variants are associated with heterogeneous diseases, even when
focusing just on changes that occur within genes.

Biological networks provide a powerful, unifying framework for
identifying disease
genes~\cite{goh2007human,barabasi2011network,cowen2017network,ozturk2018emerging}.
Genes relevant for a given disease typically target a relatively small
number of biological pathways, and since genes that take part in the
same pathway or process tend to be close to each other in 
networks~\cite{HartwellHoLe99,SpirinMi03}, disease genes  cluster
within networks~\cite{oti2007modular,gandhi2006analysis}. 
Consequently, if genes known to be causal for a particular disease are mapped onto a network, other
disease-relevant genes are likely to be found in their
vicinity~\cite{krauthammer2004molecular}.  Thus, the signal from known
disease genes can be ``propagated'' across a
network to prioritize either all genes within the network or
just candidate genes within a genomic locus where single nucleotide
polymorphisms have been correlated with an increased
susceptibility to
disease~\cite{kohler2008walking,ChenArJe2009,vanunu2010associating,navlakha2010power,pmid21699738_DADA,SmedleyKoCz14}.

While initial network approaches to identify disease genes focused on
propagating knowledge from a set of known ``gold standard'' disease
genes, with the widespread availability of cancer sequencing
data and genome-wide association studies (GWAS), the source of where
information is propagated from has shifted to genes that are newly
identified as perhaps playing a role in
disease~\cite{Netbox2010,vandin2011algorithms,lee2011prioritizing,babaei2013,jia2014varwalker,Leiserson15,CarlinFoQi19}.
For example, in the cancer context, diffusing a signal from
genes that are somatically mutated across tumors
is highly effective for identifying
cancer-relevant genes and
pathways~\cite{vandin2011algorithms,Leiserson15}; notably, while
frequency-based
approaches identify genes that ``drive'' cancer by searching for those that are
recurrently mutated across tumor samples beyond some 
background rate~\cite{LawrenceStPo13}, such a network propagation approach
can even pinpoint rarely mutated driver genes if they are within subnetworks
whose component genes, when considered together, are frequently mutated.

Thus there are two dominant network propagation paradigms 
for uncovering disease genes: spreading signal either from
well-established, annotated disease genes or from genes that have some new 
evidence of being disease-relevant.  While both have been
successful independently, we argue that both sources of information
should be utilized together, and that existing knowledge of disease genes
should inform the way new data is examined within networks.  
That is, while our prior knowledge of causal genes for a given
disease may be incomplete, it nevertheless is a valuable source of
information about the biological processes underlying the disease;
furthermore, in many cases, there is substantial prior knowledge and
there is no reason disease gene discovery should proceed \emph{de
  novo} from newly observed alterations.

In this paper, we introduce a guided network propagation framework to
uncover disease genes, where signal is propagated from new data
so as to tend to move towards genes that are closer to
known disease genes.  Our core method of propagating information
within a network is via either diffusion~\cite{qi2008finding} or
random walks with restarts (RWRs)~\cite{kohler2008walking}, as these
are mathematically sound, well-established approaches, where numerical
solutions are easily obtained.  In particular, our approach first
diffuses a signal from known disease genes, and then performs either
guided random walks or guided diffusion from the new data so as to
preferentially move towards genes that have received higher amounts of
signal from the initial set of known disease genes.  In contrast, 
previous network propagation methods for disease gene
discovery have performed diffusion or random walks uniformly from each
node (i.e., in an ``unguided'' manner, as in
e.g.,~\cite{vandin2011algorithms,jia2014varwalker}), or where the
diffusion is scaled by weights on network edges that reflect their
estimated reliabilities (e.g.,~\cite{babaei2013}). Alternatively,
several approaches have attempted to uncover disease genes by
explicitly connecting in the network genes that have genetic alterations 
 with genes that have expression
changes~\cite{KimWuPr11,bashashati2012drivernet,tiedie2013,shrestha2014hit,RuffaloKoSh15,ShiGaWa16};
while well-suited for finding genes causal for observed
expression changes, such approaches are less appropriate as a means to link prior
and new information, and our approach instead uses prior knowledge to simply
influence information propagation within the network.

We demonstrate the efficacy of our method {\tt uKIN}---\textbf{u}sing
\textbf{K}nowledge \textbf{I}n \textbf{N}etworks---by first applying
it to discover genes causal for cancer. Here, new information consists
of genes that are found to be somatically mutated in tumors---only a
small number of which are thought to play a functional role in
cancer---and prior information is comprised of subsets of ``driver''
genes known to be cancer-relevant~\cite{Futreal2004}.  In rigorous
large-scale, cross-validation style testing across 24 cancer types, we
demonstrate that propagating signal by integrating both these sources
of information performs substantially better in uncovering known
cancer genes than propagating signal from either source alone.
Notably, even using just a small number of known cancer genes (5--20)
to guide the network propagation from the set of mutated genes results
in substantial improvements over the unguided approach. Next, we
compare {\tt uKIN} to four state-of-the-art network-based methods that
use somatic mutation data for cancer gene discovery and find that {\tt
uKIN} readily outperforms them, thereby demonstrating the advantage of
additionally incorporating prior knowledge.  We also show that by using
cancer-type specific prior knowledge, {\tt uKIN} can better uncover causal
genes for specific cancer types. Finally, to showcase {\tt uKIN}'s
versatility, we show its effectiveness in identifying causal genes for
three other complex diseases, where the genes known to be associated
with the disease come from the Online Mendelian Inheritance in Man
(OMIM)~\cite{omim} and genes comprising the new information arise from
genome-wide association studies (GWAS).

%% file: methods.tex
\section*{Methods}


\begin{floatingfigure}[br]{0.7\textwidth}
\includegraphics[width=0.65\textwidth]{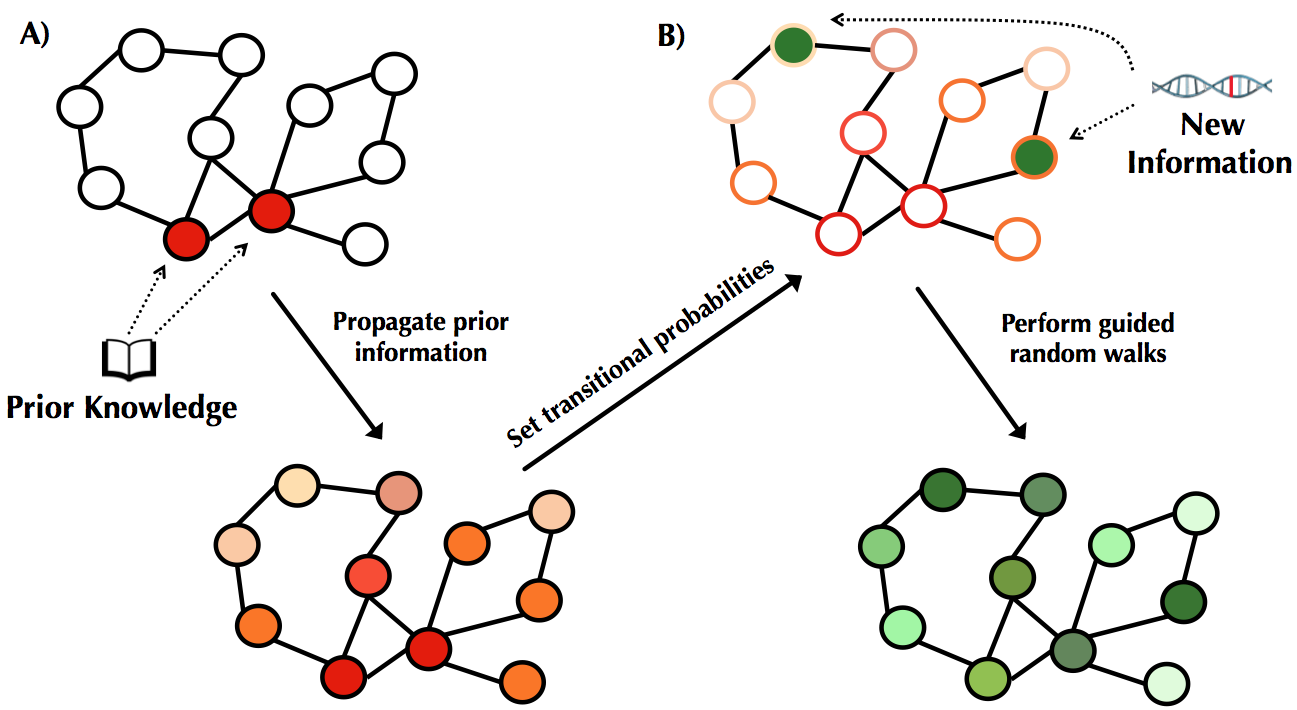}
{\caption[Overview of uKIN's algorithm]
	{\small {\bf Overview.} 
	{\bf (a)} Known
  disease-relevant genes (prior knowledge) are mapped onto an
  interaction network (shown in red, top).  Signal from this prior
  knowledge is propagated through the network via a diffusion
  approach~\cite{qi2008finding}, resulting in each gene in the network
  being associated with a score such that higher scores (visualized in
  darker shades of red, bottom) correspond to genes closer to the set of known
  disease genes.  These scores are used to set transition
  probabilities between genes such that a neighboring gene that is
  closer to the set of prior knowledge genes is more likely to be
  chosen.  {\bf (b)} Genes putatively associated with the
  disease---corresponding to the new information---are mapped onto the
  network (shown in green, top). To integrate both sources of
  information, RWRs are initiated from the set
  of putatively associated genes, and at each step, the walk either
  restarts or moves to a neighboring gene according to the transition
  probabilities (i.e., walks tend to move towards genes outlined in
  darker shades of red).  These prior-knowledge ``guided'' RWRs
  have a stationary distribution corresponding to
  how frequently each gene is visited, and this distribution is used
  to order the genes. Higher scores correspond to more frequently
  visited genes (depicted in darker greens, bottom).}
	\label{fig:ukin:fig1}}
\end{floatingfigure}
\vspace{.1in}

\noindent
{\bf Overview.} At a high level, our approach {\tt uKIN} propagates new information across a
network, while using prior information to guide this propagation (Figure~\ref{fig:ukin:fig1}).
While our approach is generally applicable, here we focus on the case
of propagating information across biological networks in order to find
disease genes. We assume that prior knowledge about a disease consists
of a set of genes already implicated as causal for that disease, and
new information consists of genes that are potentially
disease-relevant. In the scenario of uncovering cancer genes, prior
information comes from the set of known cancer genes, and new
information corresponds to those genes that are found to be
somatically mutated across patient tumors.  For other complex
diseases, new information may arise from (say) genes weakly associated
with a disease via GWAS studies or found to have \emph{de novo} or
rare mutations in a patient population of interest.


\begin{sloppypar}
The first step of our approach is to compute for each gene a measure
that captures how close it is in the network to the prior knowledge
set of genes ${\cal K}$ (Figure~\ref{fig:ukin:fig1}a). To accomplish
this, we spread the signal from the genes in ${\cal K}$ using a
diffusion kernel~\cite{qi2008finding}.  Next, we consider new
information consisting of genes ${\cal M}$ that have been identified
as potentially being associated with the disease.  As we expect those
that are actually disease-relevant to be proximal to each other and to
the previously known set of disease genes, we spread the signal from
these newly implicated genes ${\cal M}$, biasing the signal to move
towards genes that are closer to the known disease genes ${\cal K}$
(Figure~\ref{fig:ukin:fig1}b). We accomplish this by performing RWRs,
where with probability $\alpha$, the walk jumps back to one of the
genes in ${\cal M}$. That is, $\alpha$ controls the extent to which we
use new versus prior information, where higher values of $\alpha$
weigh the new information more heavily.  With probability $1-\alpha$,
the walk moves to a neighboring node, but instead of moving from one
gene to one of its neighbors uniformly at random as is typically done,
the probability instead is higher for neighbors that are closer to the
prior knowledge set of genes ${\cal K}$.  Genes that are visited more
frequently in these random walks are more likely to be relevant for
the disease because they are more likely to be part of important
pathways around ${\cal K}$ that are also close to ${\cal M}$. We thus
numerically compute the probability with which each gene is visited in
these random walks, and then use these probabilities to rank the
genes.  As an alternative to a RWR, we also experiment with
implementing the guided propagation via a diffusion
kernel~\cite{qi2008finding}.  Each step of our procedure is described
in more detail below.
\end{sloppypar}

\smallskip
\begin{sloppypar}
\noindent
{\bf Notation.} The biological network is modeled as an undirected graph
$G=(V,E)$ where each vertex represents a gene, and there is an edge
between two vertices if an interaction has been found between the
corresponding protein products. We require $G$ to be connected,
restricting ourselves to the largest connected component if necessary.
We explain our formulation with respect to cancer, but note that it is
applicable in other settings (both disease and otherwise).  The set of
genes already known to be cancer associated is denoted by ${\cal
  K}=\{k_1,k_2,...,k_l\}$.  The set of genes that have been found to
be somatically mutated in a cohort of individuals with cancer is
denoted by ${\cal M}=\{m_1,m_2,...,m_p\}$, with ${\cal F} = \{f_{m_1},
f_{m_2}, ..., f_{m_p}\}$ corresponding to the rate with which each of
these genes is mutated.  We refer to ${\cal K}$ as the prior knowledge
and ${\cal M}$ as the new information.  We assume that ${\cal
  K}\subset V$ and ${\cal M} \subset V$; in practice, we remove genes
not present in the network. The genes within ${\cal K}$ and ${\cal M}$
may overlap (i.e., it is not required that ${\cal K} \cap M =
\emptyset$).  
\end{sloppypar}


\comment{
Our method is based on the intuition that genes close to ${\cal K}$
are more likely to be involved in the same cellular processes or
pathways as genes in ${\cal K}$ and hence more likely to be relevant
for disease.  We thus perform random walks over the network $G$,
starting from ${\cal M}$ but biased towards going closer to ${\cal
  K}$, and rank genes with respect to disease relevance by how
frequently they are visited.}

\medskip
\begin{sloppypar}
 \noindent
{\bf Guided RWR Algorithm.} For each gene $i\in V$, assume that we have a measure
$q_i$ that represents how close $i$ is to the set of genes ${\cal K}$.
We will use the nonnegative vector $q$, which we describe in the next section, to
guide a random walk starting at the nodes in ${\cal M}$ and walking towards
the nodes in ${\cal K}$.  Each walk starts from a gene $i$ in ${\cal M}$,
chosen with probability proportional to its mutational rate $f_i$.  At
each step, with probability $\alpha$ the walk can restart from a gene
in ${\cal M}$, and with probability $1 - \alpha$ the walk moves to a
neighboring gene picked probabilistically based upon $q$.
Specifically, if ${\cal N}(i)$ are the neighbors of node $i$, the walk
goes from node $i$ to node $j \in {\cal N}(i)$ with probability
proportional to $q_j/\sum_{k \in {\cal N}(i)}q_k$.  That is, if at time
$t$ the walk is at node $i$, the probability that it transitions to
node $j$ at time $t+1$ is
$$p_{ij} =
(1 - \alpha) \delta_{ij}\cdot {\frac{q_j}{\sum_{k\in N(i)}q_k}} + \alpha \cdot {\frac{f_j}{\sum_{k\in
  {\cal M}}f_k}}$$
where $\delta_{ij} = 1$ if $j\in {\cal N}(i)$ 
and $0$ otherwise. Hence, the guided random walk is fully described by
a stochastic transition matrix $P$ with entries $p_{ij}$. 
By the Perron-Frobenius theorem, the corresponding random walk
has a stationary distribution $\pi$ (a left eigenvector of $P$ associated with the
eigenvalue $1$). 
If the graph $G$ is connected, then the back edges to ${\cal M}$ easily
ensure that $\pi$ is unique and can be approximated by a 
long enough random walk.
\comment{
This stochastic
matrix is non-negative and by the Perron-Frobenius theorem it has a
right eigenvector $\pi$ corresponding to eigenvalue $1$. Therefore,
$\pi P^t=\pi$ and we can efficiently compute  the stationary distribution $\pi$ that the guided
random walk converges to.
}
For each gene $i$, its score is given by the $i$th element of 
$\pi$. The genes whose nodes have high scores 
are most frequently visited and, therefore, are more likely relevant
to cancer as they are close to both the mutated starting nodes as
well as to known cancer genes. 
\end{sloppypar}

\smallskip
\noindent
{\bf Incorporating prior knowledge.}
For each gene in the network, we wish to compute how close it is to
the set of cancer-associated genes ${\cal K}$. While many approaches
have been proposed to compute ``distances'' in
networks, we use a network flow/diffusion technique where each node
$k\in {\cal K}$ introduces a continuous unitary flow which diffuses uniformly
across the edges of the graph and is lost from each node $v\in V$ in
the graph at a constant first-order rate
$\lambda$~\cite{qi2008finding}. Briefly, let $A = (a_{ij})$ denote
the adjacency matrix of $G$ (i.e., $a_{ij}=1$ if $(i, j) \in E$ and
$0$ otherwise) and let $S$ be the diagonal matrix where $s_{ii}$ is
the degree of node $i \in V$. Then, the Laplacian of the graph $G$
shifted by $\lambda$ is defined as $L=S+\lambda I - A$.  The
equilibrium distribution of fluid density on the graph is computed as
$q=L^{-1}b$~\cite{qi2008finding}, where $b$ is the 
vector with 1 for the nodes introducing the flow and 0 for the rest
(i.e., $b_i = 1$ if $v_i\in K$ and $b_i = 0$ if $v_i\notin K$ for
$\forall v_i\in V$). Note
that $L$ is diagonally dominant, hence nonsingular, for any
$\lambda\geq 0$. We  set $\lambda = 1$ in our applications.
The vector $q$ can be efficiently computed numerically.
Thus, at equilibrium, each node $i$ in the graph is associated with
a score $q_i$ which reflects how close it is to the nodes already
marked as causal for cancer.

\smallskip
\noindent
{\bf Guided diffusion.} Instead of performing RWRs to propagate
knowledge in a guided manner, it is also possible to adapt the
diffusion approach just outlined by letting $A = (a_{ij})$ be defined
such that $a_{ij} = q_j/\sum_{k \in {\cal N}(i)}q_k$, and using $A$ to compute  $L$ 
and the equilibrium density as above.

\smallskip
\noindent
{\bf Data sources and pre-processing.}  We test {\tt uKIN} on two
protein-protein interaction networks: \emph{HPRD} (Release 9\_041310)~\cite{prasad2009human} and \emph{BioGrid} (Release 3.2.99, physical
interactions only)~\cite{stark2006biogrid}.  We pre-process the
networks as in~\cite{hristov2017network}. Briefly, we remove all
proteins with an unusually high number of interactions ($>$ 900
interactions, $>$ 10 standard deviations away from the mean
number of interactions).  Additionally, to remove spurious
interactions, we remove those that have a $Z$-score normalized
diffusion state distance $>0.3$~\cite{cao2013going}. This leaves
\emph{HPRD} with 9,379 proteins and 36,638 interactions and
\emph{BioGrid} with 14,326 proteins and 102,552 interactions.

We use level 3 cancer somatic mutation data from TCGA~\cite{tcga} for 24 cancer types
(Supplemental Table~1).  For each cancer type, we process the
data as previously described and exclude samples that are
obvious outliers with respect to their total number of mutated genes~\cite{hristov2017network}. 
Our set of prior knowledge is constructed from the 719 CGC genes
that are labeled by COSMIC  (version~August 2018) as
being causally implicated in cancer~\cite{Futreal2004}. For each cancer type, our new information
consists of genes that have somatic missense or nonsense
mutations, and we compute the mutational
frequency of a gene as the number of observed somatic missense and nonsense
 mutations across tumors, divided
by the number of amino acids in the encoded protein.

We obtain 24, 28, and 63 genes associated with three complex
diseases, \emph{age-related macular degeneration (AMD)}, \emph{Amyotrophic
  lateral sclerosis (ALS)} and \emph{epilepsy}, respectively, from OMIM~\cite{omim}.  These genes are used to construct the set of
prior knowledge.  For each disease, we form the set $M$ by querying
from the GWAS database \cite{gwas} the genes implicated for the
disease and using the corresponding $p$-values to compute the starting
frequencies $f$. Specifically, for each disease, for each GWAS study
$i$, if a gene $j$'s $p$-value is $p_{i,j}$, we set its frequency to
$\log(p_{i,j})/\sum_k\log(p_{i,k})$ and then for each gene average
these frequencies over the studies.

\smallskip
\noindent
{\bf Performance evaluation.}  To evaluate our method in the context
of cancer, we subdivide the CGC genes that appear in our network into
two subsets. We randomly draw from the CGCs 400 genes to form a set
${\cal H}$ of positives that we aim to uncover. From the remaining 199
CGCs present in the network, we randomly draw a fixed number $l$ to
represent the prior knowledge ${\cal K}$ and run our framework.  As we
consider an increasing number of most highly ranked genes, we compute
the fraction that are in the set ${\cal H}$ of positives. All CGC
genes not in ${\cal H}$ are ignored in these calculations.
Importantly, the genes in ${\cal K}$ which are used to guide the
network propagation are never used to evaluate the performance of {\tt
uKIN}. Note that this testing set up, which measures
performance on ${\cal H}$, allows us to compare performance of {\tt
uKIN} when choosing prior knowledge sets of different size $l$ from
the CGC genes not in ${\cal H}$.

We also compute area under the precision-recall curves (AUPRCs).  In
this case, all CGC genes in ${\cal H}$ are considered positives, all CGC
genes not in ${\cal H}$ are neutral (ignored), and all other genes are
negatives. Though we expect that there are genes other than those
already in the CGC that play a role in cancer, this is a standard
approach to judge performance (e.g., see~\cite{jia2014varwalker}) as
cancer genes should be highly ranked. To focus on performance with respect to the top predictions,  we compute AUPRCs using the
top 100 predicted genes. To better estimate AUPRCs and account for the randomness in sampling, we
repeatedly draw (10 times) the set ${\cal H}$ and for each draw we sample the
genes comprising the prior knowledge ${\cal K}$ 10 times. The final AUPRC
results from averaging the AUPRCs across all 100 runs.
 
We compare {uKIN} on the cancer datasets to the frequency-based method
{\tt MutSigCV 2.0}~\cite{lawrence2013mutational} and four
network-based methods, {\tt DriverNet}~\cite{bashashati2012drivernet},
{\tt Muffinn}~\cite{cho2016muffinn}, {\tt
nCOP}\cite{hristov2017network} and {\tt HotNet2}~\cite{Leiserson15}.
All methods are run on each of the 24 cancer types with their default
parameters.  {\tt Muffinn}, {\tt nCOP} and {\tt HotNet2} are run on
the same network as {\tt uKIN}, whereas MutSigCV
does not use a network and DriverNet instead uses an influence (i.e.,
functional interaction) graph and transcriptomic data (we use their
default influence graph and provide as input TCGA normalized
expression data).  Since {\tt uKIN} uses a subset of CGCs as prior
knowledge, we ensure that all methods are evaluated with respect to
the hidden sets ${\cal H}$ (i.e., of CGCs not used by {\tt
uKIN}).  Though we could just consider performance
with respect to one hidden set, considering multiple sets enables a
better estimate of overall performance.   For these comparisons,
{\tt uKIN} with $\alpha = 0.5$ is run 100 times, as
described above, with 20 randomly sampled genes comprising the prior
knowledge, and evaluation is performed with respect to the genes in
the hidden sets.  All methods' AUPRCs are computed using the same
randomly sampled test sets ${\cal H}$ and averaged at the end.  Since
{\tt HotNet2} outputs a set of predicted cancer-relevant genes and
does not rank them, we cannot compute AUPRCs for it; instead we
compute precision and recall for its output with respect to the test
sets ${\cal H}$ and compare to {\tt uKIN}'s when considering the same
number of top scoring genes.

To evaluate our method in the context of the three complex diseases, we subdivide
evenly the set of OMIM genes associated with each disease into the 
prior knowledge set ${\cal K}$ and the set of positives ${\cal H}$.
As with the cancer data, we do this repeatedly (100 times) and average
AUPRCs at the end.

%% file: results.tex
\medskip
\section*{Results}

We first apply our method {\tt uKIN} to uncover cancer genes.  Genes
that have missense and nonsense somatic mutations comprise the new
information, and random walks start from these genes with probability
proportional to their mutation rates.  We apply our approach to data from 24
cancer types, but showcase results for glioblastoma multiforme (GBM).
All results in the main paper use the \emph{HPRD} protein-protein
interaction network~\cite{prasad2009human}, with results shown for
\emph{BioGrid}~\cite{stark2006biogrid} in the Supplement.

\smallskip
\noindent
{\bf uKIN successfully integrates prior knowledge and new
  information.}  We compare {\tt uKIN}'s performance when using both prior and new
knowledge (RWRs with $\alpha=0.5$), to versions of {\tt uKIN} using either
only new information ($\alpha = 1$) or only prior information ($\alpha
= 0$).  Briefly, we use $20$ randomly drawn
CGCs to represent the prior knowledge ${\cal K}$ and another $400$
randomly drawn CGCs to be the hidden set $H$ of unknown
cancer-relevant genes that we aim to uncover (see {\bf Performance
  evaluation} for details).  We repeat this process 100 times, each
time spreading signal using the diffusion
approach~\cite{qi2008finding} before performing RWRs from the genes
observed to be somatically mutated.  For each run, we analyze the
ranked list of genes output by {\tt uKIN} as we consider an increasing
number of output genes, and average across runs  the fraction
that are members of the hidden set ${\cal H}$ consisting of cancer
driver genes.

\begin{figure}[t!]
\includegraphics[width=\textwidth]{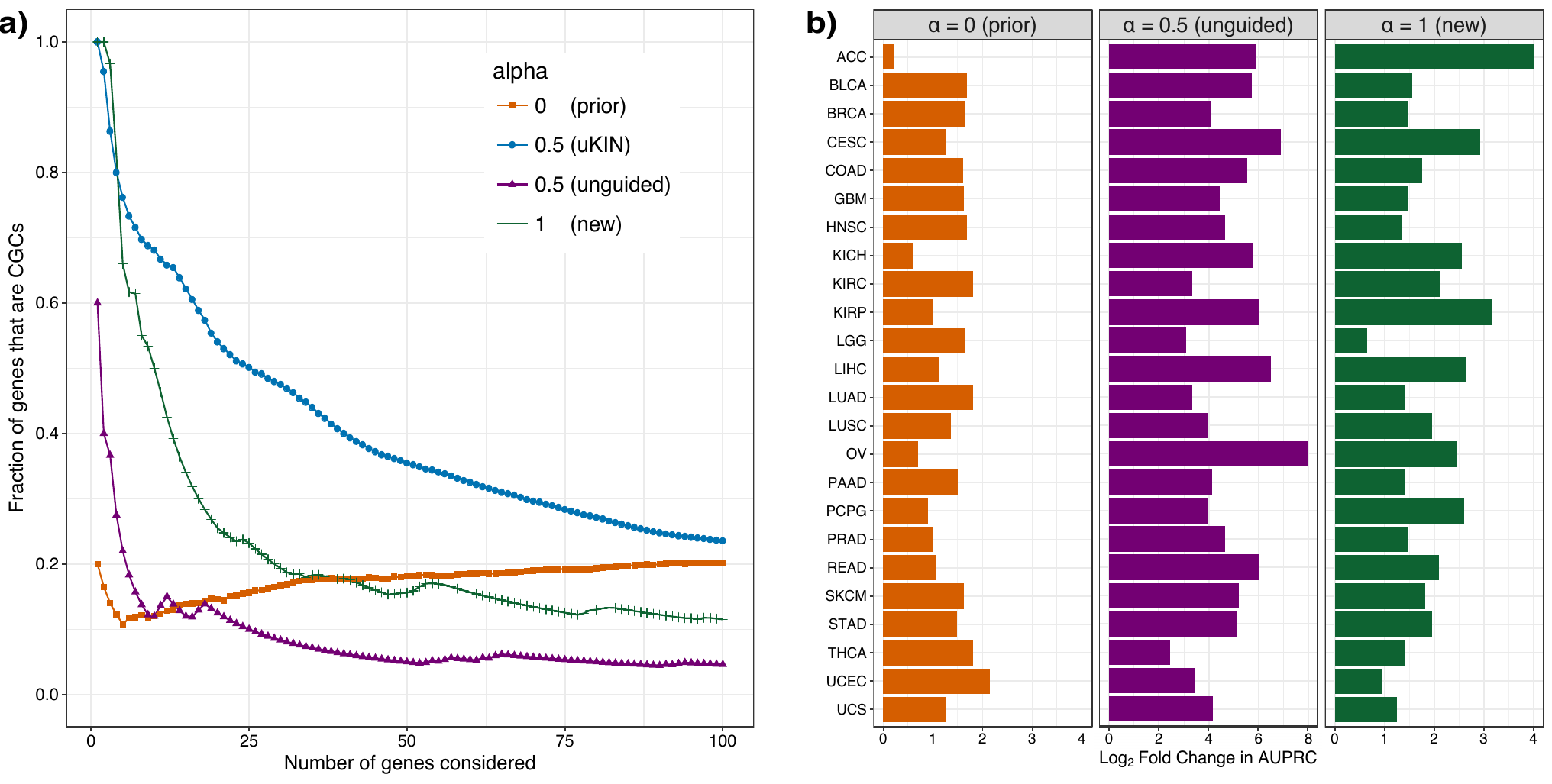}
\caption[uKIN successfully integrates new information and prior
  knowledge] { \small {\bf {\tt uKIN} successfully integrates new
    information and prior knowledge.}  {\bf (a)} We illustrate the
  effectiveness of our approach {\tt uKIN} on the GBM data set and the
  HPRD protein-protein interaction network using $20$ randomly drawn
  CGCs to represent the prior knowledge.  We combine prior and new
  knowledge using a restart probability of $\alpha = 0.5$ (blue line).
  As we consider an increasing number of high scoring genes, we plot
  the fraction of these that are part of the hidden set of CGCs.  As
  baseline comparisons, we also consider versions of our approach
  where we use only the new information ($\alpha=1$) and order genes
  by their mutational frequency (green line); where we use new information
  to perform \emph{unguided} random walks with $\alpha = 0.5$ and
  order genes by their probabilities in the stationary distribution of
  the walk (which uses new information but not prior information,
  purple line); and where we use only prior information ($\alpha=0$)
  and order genes based on  information propagated from the set of
  genes comprising our prior knowledge (orange line).  Integrating
  both prior and new sources of information results in better
  performance.  {\bf (b)} The performance of {\tt uKIN} when
  integrating information at $\alpha = 0.5$ is compared to the three
  baseline cases where either only prior information is used ($\alpha
  = 0$, left) or when only new information is used ($\alpha = 1$,  right and unguided RWRs with $\alpha = 0.5$, middle).  In all three
  panels, for each cancer type, we plot the $\log_2$ ratio of the AUPRC of {\tt
    uKIN} with guided RWRs with $\alpha = 0.5$ to the  AUPRC of the other approach. Across all 24 cancer
  types, using both sources of information outperforms using just one
  source of information. }
	\label{fig:ukin:fig2}
\end{figure}

For $\alpha = 0.5$, we observe that a large fraction
of the top predicted genes using the GBM dataset are part of the hidden set of known cancer
genes (Figure~\ref{fig:ukin:fig2}a).  
At $\alpha = 1$, our method completely ignores both the network and
the prior information ${\cal K}$ and is equivalent to ordering the
genes by their mutational frequencies.  The
very top of the list output by {\tt uKIN} when $\alpha = 1$ consists
of the most frequently mutated genes (in the case of GBM,
this includes \emph{TP53} and \emph{PTEN}).  As we
consider an increasing number of genes, ordering them by mutational
frequency is clearly outperformed by {\tt uKIN} with $\alpha=0.5$.
At the other extreme with $\alpha=0$, the starting locations and their
mutational frequencies are ignored as the random walk is memoryless
and the stationary distribution depends only upon the propagated prior
information $q$.  As expected, performance is considerably worse than
when running {\tt uKIN} with $\alpha = 0.5$.  Nevertheless, we observe that several
CCGs are found for $\alpha=0$; this is due to the fact that known
cancer genes tend to cluster together in the network~\cite{Netbox2010} and
our propagation technique ranks highly the genes close to the genes in
${\cal K}$.


We also consider {\tt uKIN}'s performance as compared to an \emph{unguided} walk with
the same restart probability $\alpha = 0.5$. In this case, the walk
selects a neighboring node to move to uniformly at random. The
stationary distribution that the walk converges to depends upon the
starting locations and the network topology but is independent of the
prior information.  Such a walk provides a good baseline to judge the
impact the propagated prior information $q$ has on the performance of
our algorithm, and is an approach that has been widely
applied~\cite{kohler2008walking}.  As evident in
Figure~\ref{fig:ukin:fig2}a, an \emph{unguided} walk (purple line) performs 
considerably worse than {\tt uKIN} with $\alpha = 0.5$, highlighting the importance of $q$ in \emph{guiding}
the walk.

Notably, the trends we observe on GBM hold across all 24 cancers
(Figure~\ref{fig:ukin:fig2}b).  For each cancer type, we consider the
$\log_2$ ratio of the AUPRC of the version of {\tt uKIN} that uses both prior and new
information with $\alpha=0.5$ to the AUPRC for each of the other
variants.  For all cancer 24 cancers, when {\tt uKIN} 
uses both prior and new information with $\alpha = 0.5$, it outperforms
the cases when using only prior information (Figure~\ref{fig:ukin:fig2}b, left) or
using only new information (Figure~\ref{fig:ukin:fig2}b, middle and right).
Further, we observe this improvement when using both prior and
new information across all cancers for a wide range of $\alpha$ ($0.2
< \alpha < 0.8$, data not shown), clearly demonstrating that using
both sources of information is beneficial.


\smallskip
\noindent
{\bf uKIN is effective in uncovering cancer-relevant genes.}  We next
evaluate {\tt uKIN}'s performance in uncovering cancer-relevant genes
as compared to several previous methods.  These methods do not use any
prior knowledge of cancer genes, and any performance differences
between {\tt uKIN} and them may be due either to the use of this
important additional source of information or to specific algorithmic
differences between the methods. Nevertheless, such comparisons are
necessary to get an idea of how well {\tt uKIN} performs as compared
to the current state-of-the-art.  All methods are run and AUPRCs
computed as described in {\bf Methods}.  First, we compare uKIN with
$\alpha = 0.5$ to {\tt MutSigCV~2.0}~\cite{lawrence2013mutational},
perhaps the most widely used frequency-based approach to identify
cancer driver genes.  We find that {\tt uKIN} outperforms {\tt
MutSigCV~2.0} on 22 of 24 cancer types
(Figure~\ref{fig:ukin:fig3}a). Next, we compare {\tt uKIN} to three
network-based approaches (Figure~\ref{fig:ukin:fig3}b): {\tt Muffinn}
\cite{cho2016muffinn}, which considers mutations found in
interacting genes; {\tt DriverNet} \cite{bashashati2012drivernet}, 
which finds driver genes by uncovering sets of somatically
mutated genes that are linked to dysregulated genes; and {\tt nCOP}
\cite{hristov2017network}, which examines the
per-individual mutational profiles of cancer patients in a biological
network. {\tt uKIN} exhibits superior
performance across all cancer types when compared to {\tt DriverNet}, and
outperforms {\tt Muffinn} in 23 out of 24 cancer types and {\tt nCOP}
in 17 of the 24 cancer types.  In many cases,
  the performance improvements of {\tt uKIN} are substantial (e.g.,
  more than a 2-fold improvement for 12, 10, 3 and 4 cancer types for
  {\tt MutSigCV}, {\tt DriverNet}, {\tt Muffin} and {\tt nCOP},
  respectively).  We also compare to {\tt Hotnet2}
\cite{Leiserson15}, whose core algorithmic component is
diffusion~\cite{qi2008finding}, and as such {\tt uKIN} is more similar
to it than other methods.  {\tt Hotnet2} does not output a ranked list
of genes, so we instead examine the list of genes highlighted by both
methods. We find that {\tt uKIN} exhibits higher precision and recall
than {\tt Hotnet2} for all cancer types (Suppl.\
Figure~\ref{fig:ukin_suppl:sfig_ukin_hotnet}); since both {\tt uKIN}
and {\tt Hotnet2} are network propagation approaches, these
performance improvements illustrate the benefit of using prior
information in identifying cancer-relevant genes.


\begin{floatingfigure}[br]{0.7\textwidth}
\includegraphics[width=0.65\textwidth]{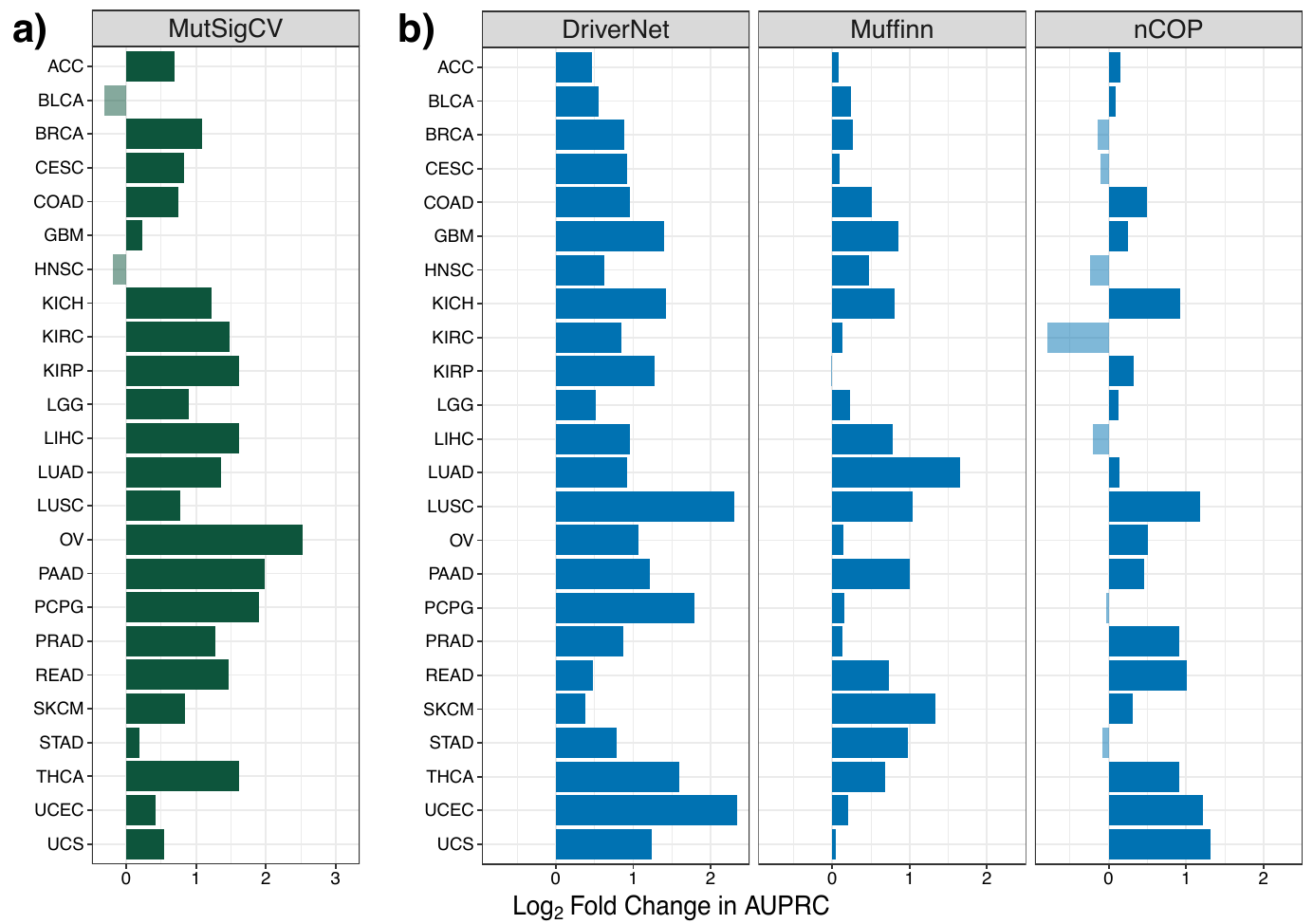}
\caption[uKIN is more effective than other methods
  in identifying known cancer genes]
	{ \small  {\bf {\tt uKIN} is more effective than other methods
  in identifying known cancer genes.}  For each method, for each cancer type, we plot the $\log_2$
  ratio of {\tt uKIN}'s AUPRC to its AUPRC.  {\bf (a)}
  Comparison of {\tt uKIN} to {\tt MutSigCV~2.0}, a state-of-the-art frequency-based
  approach. {\tt uKIN} outperforms {\tt MutSigCV~2.0} on 22 of the 24
  cancer types.  {\bf (b)} Comparison of {\tt uKIN} to  {\tt DriverNet} (left),
  {\tt Muffinn} (middle),  and {\tt
  nCOP} (right).  Our approach {\tt
  uKIN} outperforms {\tt DriverNet} on all cancer types, 
	{\tt Muffinn} on all but one cancer type and {\tt nCOP} 	on 17 out of 24 cancer types.}
	\label{fig:ukin:fig3}
\end{floatingfigure}

\comment{
\begin{sloppypar}
In several cancers, the performance improvements of {\tt uKIN} are
substantial. For example, {\tt uKIN} has a four-fold improvement
over {\tt MutSigCV~2.0} in predicting cancer genes for ovarian cancer
(OV) and pancreas adenocarcinoma (PAAD), \textcolor{red}{and a four-fold improvement
over {\tt DriverNet} for uterine} corpus endometrial carcinoma (UCEC)
and lung squamous cell carcinoma (LUSC).  The limited number of
patient samples available for uterine carcinosarcoma (UCS) limits
{\tt nCOP}'s perfomance \cite{hristov2017network} whereas {\tt uKIN} is
able to leverage the prior knowledge available, resulting in 
{\tt uKIN}'s two fold improvement over {\tt nCOP}; this
highlights the benefits from incorporating existing knowledge
of disease-relevant genes, especially when the new data is sparse. 
\end{sloppypar}
}

\comment{    We next demonstrate that {\tt uKIN}
    highlights genes with a broad range of mutational rates.  When we
    run {\tt uKIN} on each of the 24 cancer types using prior
    knowledge consisting of 20 genes sampled 100 times, and consider
    the top $100$ predictions (averaged across the runs), we observe
    that these genes have vastly diverse mutational rates
    (Suppl.\ Figure~\ref{fig:ukin_suppl:sfig_violin_all_c} for
    all cancer types).
}

\smallskip
\begin{sloppypar}
\noindent
{\bf Robustness tests.}  The overall results shown hold when we use
different lists of known cancer genes as a gold standard
(Suppl.\ Figure~\ref{fig:ukin_suppl:sfig2}a), different numbers
of predictions considered when computing AUPRCs (Suppl.\
Figure~\ref{fig:ukin_suppl:sfig2}b), and different networks
(Suppl.\ Figure~\ref{fig:ukin_suppl:sfig2}c). Further, we
confirm the importance of network structure to {\tt uKIN}, by running
{\tt uKIN} on two types of randomized networks, degree-preserving and
label shuffling, and show that, as expected, overall performance
deteriorates across the cancer types (Suppl.\ 
Figure~\ref{fig:ukin_suppl:sfig2}d); we note that while network
structure is destroyed by these randomizations, per-gene mutational
information is preserved, and thus highly mutated genes are still
output.
\end{sloppypar}

We also determine the effect of the amount of prior knowledge
for {\tt uKIN}, and find that while performance increases with larger
numbers of genes comprising our prior knowledge, even as few as five
prior knowledge genes leads to a $\sim$$4$-fold improvement over ranking genes
by mutational frequency (Suppl.\ Figure~\ref{fig:ukin:fig5}a). Finally, we
investigate the effect of some incorrect prior knowledge, and find
that while {\tt uKIN}'s performance decreases with more incorrect
knowledge, {\tt uKIN} with $\alpha = 0.5$ performs reasonably with
$<20\%$ incorrect annotations (Suppl.\ Figure~\ref{fig:ukin:fig5}b).

\smallskip
\noindent
{\bf Alternate formulations.} We also tested guided diffusion from the
    somatically mutated genes instead of RWRs (see {\bf Methods}). We
    empirically find that, for $\alpha = 0.5$, diffusion with
    $\lambda=1$ yields nearly identical per-gene scores on the cancer
    datasets we tested (GBM and kidney renal cell carcinoma).
    Similarly, for other $\alpha$, we were able to find values of
    $\lambda$ such that the RWRs and diffusion have highly similar
    results.  On the other hand, replacing the initial diffusion from
    the prior knowledge with a RWR (with $\alpha$ = 0.5) results in
    somewhat worse performance (e.g., $\sim$$10\%$ drop in AUPRC for
    GBM). 
    \comment{
Overall, the per-node values resulting from the initial
    diffusion from prior information with $\lambda$ = 1 tend to have
    high correlations with the stationary distributions resulting from
    RWRs (with $\alpha = 0.5$) from the prior information (Pearson's correlations $>$ .7 and Spearman's correlations $>$ .9).
Finally, we also experimented
    with a single phase RWR approach where with probability~$\alpha$
    the walk restarts at either the new data ${\cal M}$ or the prior
    knowledge ${\cal K}$ (the two choices traded off with probability
    parameter $\beta$); in preliminary testing, this approach performs
    notably worse than the described approach (details omitted due to
    lack of space).}

\smallskip
\begin{sloppypar}
\noindent
 {\bf uKIN highlights infrequently mutated cancer-relevant genes.}  A
 major advantage of network-based methods is that they are able to
 identify cancer-relevant genes that are not necessarily mutated in
 large numbers of patients~\cite{Leiserson15}.  We
 next analyze the mutation frequency of genes output by {\tt uKIN}
 with $\alpha = 0.5$.  In particular, for each cancer type, for each
 gene, we obtain a final score by averaging scores across the 100 runs
 of {\tt uKIN}; to prevent ``leakage'' from the prior knowledge set,
 if a gene is in the set of prior knowledge genes ${\cal K}$ for a
 run, this run is not used when determining its final score.   We confirm
 that, for all cancer types, the top scoring genes 
 exhibit diverse mutational rates, and include both frequently
 and infrequently mutated genes (Suppl.\
 Figure~\ref{fig:ukin_suppl:sfig_violin_all_c}).
 \end{sloppypar}

We next highlight some infrequently mutated genes in GBM that are given 
high final scores by {\tt uKIN}  (i.e., are predicted as cancer-relevant).  For example,
\emph{LAND1A} and \emph{SMAD4} are two well known cancer players
that are highly ranked by {\tt uKIN}, and that have mutational rates
in GBM that are in the bottom 70\% of all genes and are therefore hard
to detect with frequency-based approaches.  Of {\tt uKIN}'s top 100
scoring genes, 23 are are in the bottom half with respect to
mutational rates, and 5 of these are CGCs ($p < 10^{-2}$,
hypergeometric test).  When considering the top scoring 100 genes by
{\tt uKIN} for each cancer type, those that have mutational ranks in
the bottom half of all genes are each found to have a statistically
significant enrichments of CGC genes.  Thus, {\tt uKIN} provides a
means for pulling out cancer genes from the ``long
tail''~\cite{Garraway2013} of infrequently mutated genes.


\begin{floatingfigure}[htbr]{0.7\textwidth}
\includegraphics[width=0.65\textwidth]{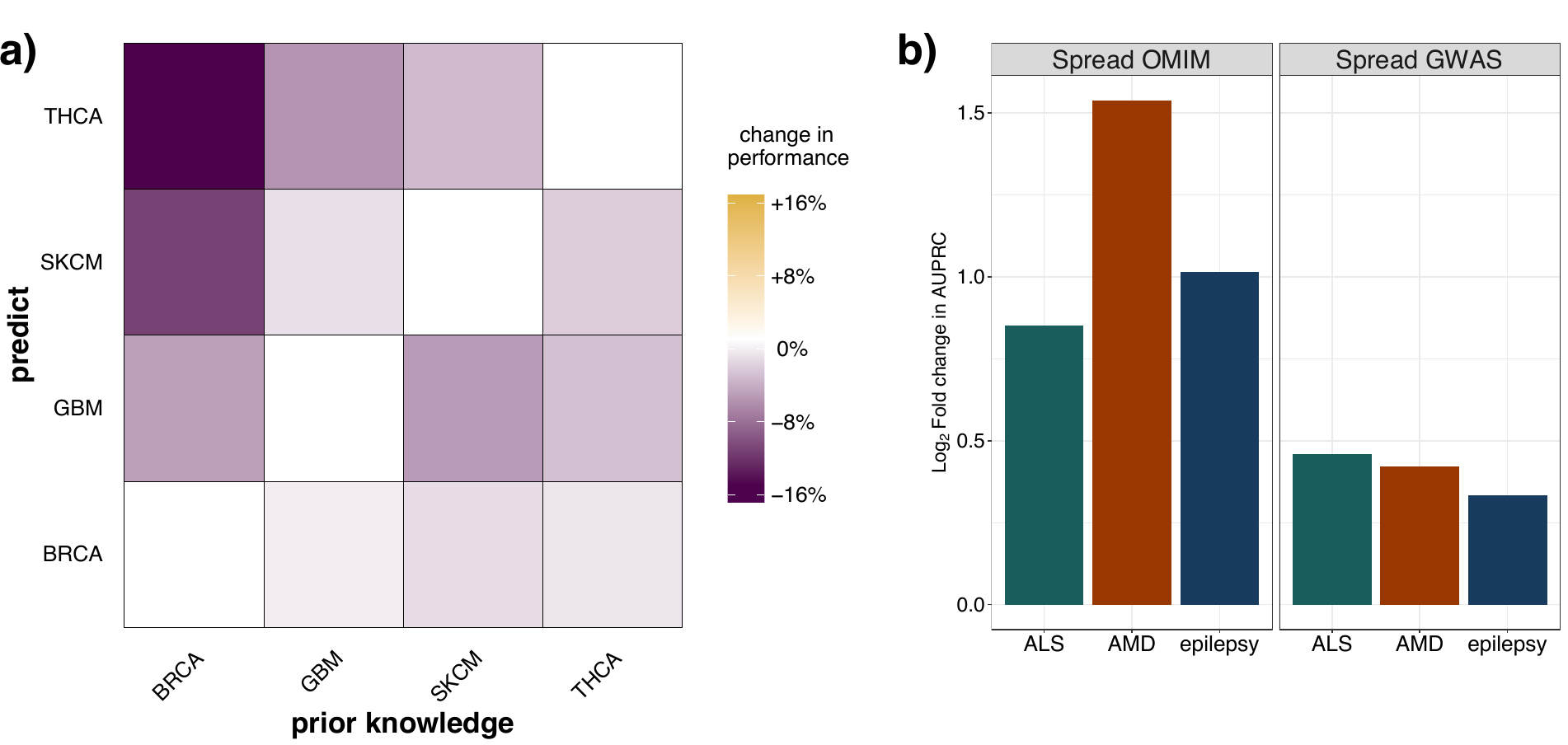}
\caption[Cancer-specific knowledge yields better performance]
{\small {\bf (a) Use of cancer-type specific knowledge improves
    performance.} For four cancer types, BRCA, GBM, SKCM, and THCA, we
    consider the performance of {\tt uKIN} with $\alpha=0.5$ when
    using TCGA mutational data for that cancer type with prior
    knowledge consisting of genes known to be driver in that cancer
    type, as compared to performance when the prior knowledge set
    consists of genes that are annotated as driver only for one of the
    other three cancer types.  For each cancer, performance is
    measured by the average ranking by {\tt uKIN} of genes known to be
    driver for that cancer.  For all combinations of possible prior
    knowledge sets ($x$-axis) and specific cancer gene sets that we
    wish to recover ($y$-axis), using prior knowledge from another
    cancer (off diagonal entries) leads to a decrease in performance
    as compared to the corresponding pairs (diagonal entries), as
    measured by the increase in {\tt uKIN}'s average ranking of genes
    we aimed to uncover.  {\bf (b) {\tt uKIN} is effective in
    identifying complex disease genes.}  We demonstrate the
    versatility of the {\tt uKIN} framework by integrating OMIM and
    GWAS data for three complex diseases, \emph{ALS}, \emph{epilepsy}
    and \emph{AMD}.  For each disease, we compare {\tt uKIN}'s
    performance when using OMIM annotated genes as prior information
    and GWAS hits as new information with $\alpha=0.5$, to baseline
    versions that propagate only information via diffusion from OMIM
    (left) or GWAS studies (right).  In all cases, we plot the
    $\log_2$ ratio of the AUPRC obtained by {\tt uKIN} using both
    prior and new information to the baseline
    methods.}  \label{fig:ukin:fig4}
\end{floatingfigure}

\comment{
To assess the ability of our method to leverage
    knowledge specific to different cancer types, we split the genes
    from CGC annotated to be drivers for a particular cancer type into
    two sets. We use the first one as prior knowledge while trying to
    uncover the genes in the second. This process is repeated 100
    times.  Next, we use the set of genes belonging to a different
    cancer type as prior knowledge while still trying to uncover the
    genes in the original cancer of interest.  \textcolor{red}{As compared to the
    correct pairs (diagonal entries), this leads to a decrease in
    performance (as measured by the increase in {\tt uKIN}'s average
    ranking of genes we aimed to uncover) across all possible prior
    knowledge ($x$-axis) and specific cancer gene sets ($y$-axis)
    combinations.} 
}

\begin{sloppypar}
In addition to highlighting known cancer genes, {\tt uKIN} also ranks
highly several non-CGC genes that may or may not play a functional role in cancer, as our knowledge of cancer-related genes is
incomplete. Among these novel predictions for GBM are \emph{ATXN1}, 
\emph{SMURF1}, and \emph{CCR3}, all of which have been recently
suggested to play a role in cancers~\cite{kang2017ataxin,li2017smurf1,lee2016crosstalk}
and are each mutated in less than 5\% of the samples.
\emph{ATXN1} is a chromatin-binding factor that plays a critical role 
in the development of spinocerebellar ataxia, a neurodegenerative 
disorder~\cite{rousseaux2018atxn1}, and mutants of \emph{ATXN1} have been found 
to stimulate the proliferation of cerebellar 
stem cells in mice~\cite{edamakanti2018mutant}. This is a promising 
gene for further investigation because glioblastoma is a cancer that 
usually starts in the cerebrum and the potential role of \emph{ATXN1} 
in tumorigenesis has only recently been suggested~\cite{kang2017ataxin}. 
\emph{SMURF1} and its highly ranked by {\tt uKIN} network-interactor \emph{SMAD1} 
have already been implicated in the development of several 
cancers~\cite{yang2017smad1}. \emph{SMURF1} also interacts with the 
nuclear receptor \emph{TLX} whose inhibitory role in 
glioblastoma has been revealed~\cite{johansson2016nuclear}.
Overall, we also find that the top scoring genes by {\tt uKIN} for GBM are
enriched in many KEGG pathways and GO terms relevant for cancer,
including \emph{microRNAs in cancer}, \emph{cell proliferation},
\emph{choline metabolism in cancer} and \emph{apoptosis}
(Bonferroni-corrected p $<$ 0.001, hypergeometric test).
\end{sloppypar}

\smallskip
\begin{sloppypar}
\noindent
{\bf Cancer-type specific prior knowledge yields better performance.}  In
several cases, CGC genes are annotated with the specific cancers they
play driver roles in. We next test how {\tt uKIN}'s performance
changes when using such highly specific prior knowledge. We consider
four cancer types, GBM, breast invasive carcinoma (BRCA), skin
cutaneous carcinoma (SKCM), and thyroid carcinoma (THCA), with 33, 32,
42 and 29 CGC genes annotated to them, respectively.  We repeatedly
split each of these sets of genes in half, and use half as the set
${\cal K}$ of prior knowledge, and the other half as the set ${\cal
H}$ to test performance.
\end{sloppypar}

\begin{sloppypar}
We first use knowledge consisting of genes specific to a cancer type of interest together with the TCGA data for that cancer to uncover that cancer's specific 
drivers. Given the small number of genes annotated to each
cancer, we assess performance by, for each of these
    genes, computing the rank of its score by {\tt uKIN} over the splits where these genes are in 
${\cal H}$.  Next, for the \emph{same} cancer type, we use a set ${\cal
  K}$ corresponding to a \emph{different} cancer type as prior
knowledge (excluding any genes that are annotated to the \emph{original}
cancer type) while still trying to uncover the genes in the
\emph{original} cancer of interest (i.e., using TCGA mutational data and ${\cal
  H}$ belonging to the \emph{original} cancer type). That is, we are
testing the performance of {\tt uKIN} when using knowledge
corresponding to a different cancer type.  For all four cancer types,
we find that performance is best when {\tt uKIN} uses prior
knowledge for the same cancer cancer type (Figure~\ref{fig:ukin:fig4}a), as
genes in ${\cal H}$ appear higher in the list of genes output by
{\tt uKIN}.  This suggests that {\tt uKIN} can utilize cancer-type
specific knowledge and highlights the benefits of having accurate
prior information.
\end{sloppypar}

\smallskip
\begin{sloppypar}
\noindent
 {\bf Application to identify disease genes for complex inherited
   disorders.}  A major advantage of our method is that it can be
   easily applied in diverse settings. As proof of concept, we apply
   {\tt uKIN} to detect disease genes for three complex
   diseases: \emph{AMD}, \emph{ALS} and \emph{epilepsy}. For each
   disease, we randomly split in half the OMIM database's \cite{omim}
   list of genes associated with the disease 100 times to form the set
   of prior knowledge ${\cal K}$ and the hidden set ${\cal H}$. We use
   the GWAS catalogue list of genes with their corresponding
   $p$-values to form the set ${\cal M}$. For all three diseases, {\tt
   uKIN} combining both GWAS and OMIM sources of information
   ($\alpha=0.5$) performs better than diffusing the signal with $\lambda = 1$ using only
   knowledge from OMIM (Figure~\ref{fig:ukin:fig4}b, left panel).  For
   each of these diseases, there is virtually no overlap between the
   GWAS hits ${\cal M}$ and a set of OMIM genes ${\cal H}$; simply
   sorting genes by their significance in GWAS studies (i.e., {\tt
   uKIN} with $\alpha = 1$) results in AUPRC of 0.  Instead, we spread
   information from the set of GWAS genes ${\cal M}$ in the same
   fashion as from OMIM and observe again that using this single
   source of information alone does not work as well as {\tt uKIN}'s using both GWAS and OMIM information together    (Figure~\ref{fig:ukin:fig4}b, right panel).
\end{sloppypar}

%% file: discussion.tex
\section*{Discussion}

In this paper, we have shown that {\tt uKIN}, a network propagation
method that incorporates both existing knowledge as well as new
information, is a highly effective and versatile approach for
uncovering disease genes.  Our method is based upon the intuition that
prior knowledge of disease-relevant genes can be used to guide the way
information from new data is spread and interpreted in the context of
biological networks. Because {\tt uKIN} uses prior knowledge, it has
higher precision than other state-of-the-art methods in detecting
known cancer genes.  Further, it excels at highlighting infrequently
mutated genes that are nevertheless relevant for cancer.
Additionally, we have shown that {\tt uKIN} can be applied to discover
genes relevant for other complex diseases as well.

The framework presented here can be extended in a number of natural
ways.  First, in addition to positive knowledge of known disease
genes, we may also have ``negative'' knowledge of genes that are not
involved in the development of a given disease.  These genes can
propagate their ``negative'' information, thereby biasing the random
walk to move away from their respective modules and perhaps further
enhancing the performance of our method.  Second, {\tt uKIN} is likely
to benefit from incorporating edge weights that reflect the
reliability of interactions between proteins; these weights will have
an impact on both the propagation of prior knowledge as well as the
guided random walks.  Third, since a recent
study~\cite{przytycki2017differential} has shown that contrasting
cancer mutation data with natural germline variation data helps boost
the true disease signal by downgrading genes that vary frequently in
nature, {\tt uKIN}'s performance may benefit from scaling the starting
probabilities of the new putatively implicated genes to account for
their variation in healthy populations.  Fourth, while here we have
demonstrated how {\tt uKIN} can use cancer-type specific knowledge,
cancers of the same type can often be grouped into distinct subtypes,
and such highly-detailed knowledge may improve {\tt uKIN}'s
performance even further.  Finally, we note that network propagation
approaches have been applied to non-disease settings as well,
including biological process
prediction~\cite{WangMa11,NabievaJiAg05}. We conjecture that our
guided network propagation approach will additionally be useful in
other scenarios in computational biology, including where new data
(e.g., arising from functional genomics screens) need to be
interpreted in the context of what is already known about a biological
process of interest.

In conclusion, {\tt uKIN} is a flexible and effective method that
handles diverse types of new information. As our knowledge of
disease-associated genes continues to grow and be refined, and as new
experimental data becomes more abundant, we expect that the power of {\tt uKIN}
for accurately prioritizing disease genes will continue to increase.

%% file: supplement.tex
\newpage
\beginsupplement

\noindent
{\large
\section*{Supplementary Figures and Tables}
}

\bigskip

\noindent
The following pages contain 1 table and 4 supplementary figures that support the findings of the
main paper.

\clearpage
\newpage

\begin{table}[t!]
\includegraphics[width=1\textwidth]{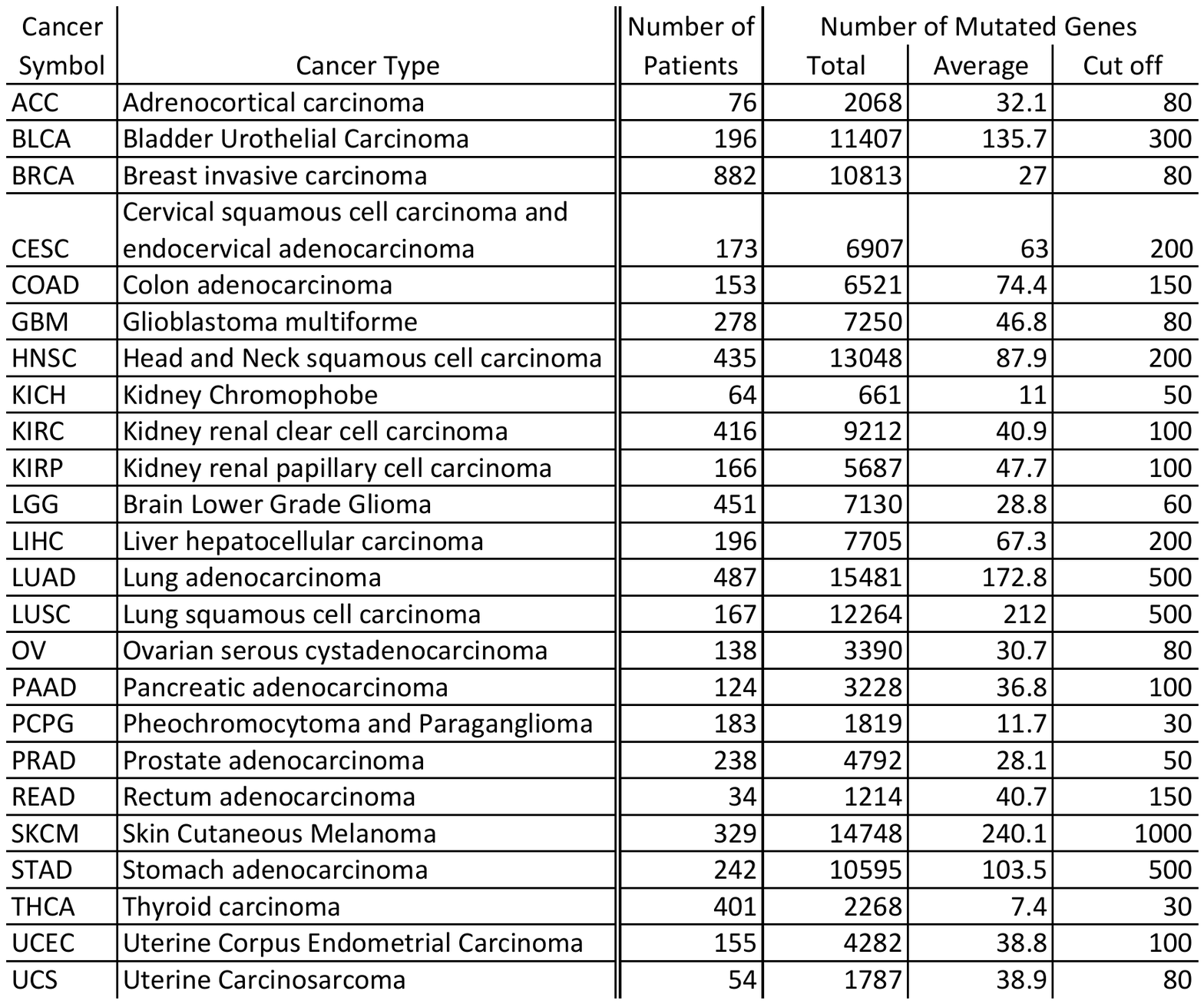}
\vspace{-3in}
\caption{{\bf TCGA dataset and statistics.} We list the 24 cancer types studied
  along with their abbreviations. For each cancer type, we give the
  total number of patient samples considered after highly mutated
  samples are filtered out, the total number of mutated genes across
  these samples, the average number of mutated genes across
  all samples, and the cutoff on the number of mutated genes
  within a sample that was used to filter samples.}
\end{table}


\begin{figure}[t]
\includegraphics[width=0.8\textwidth]{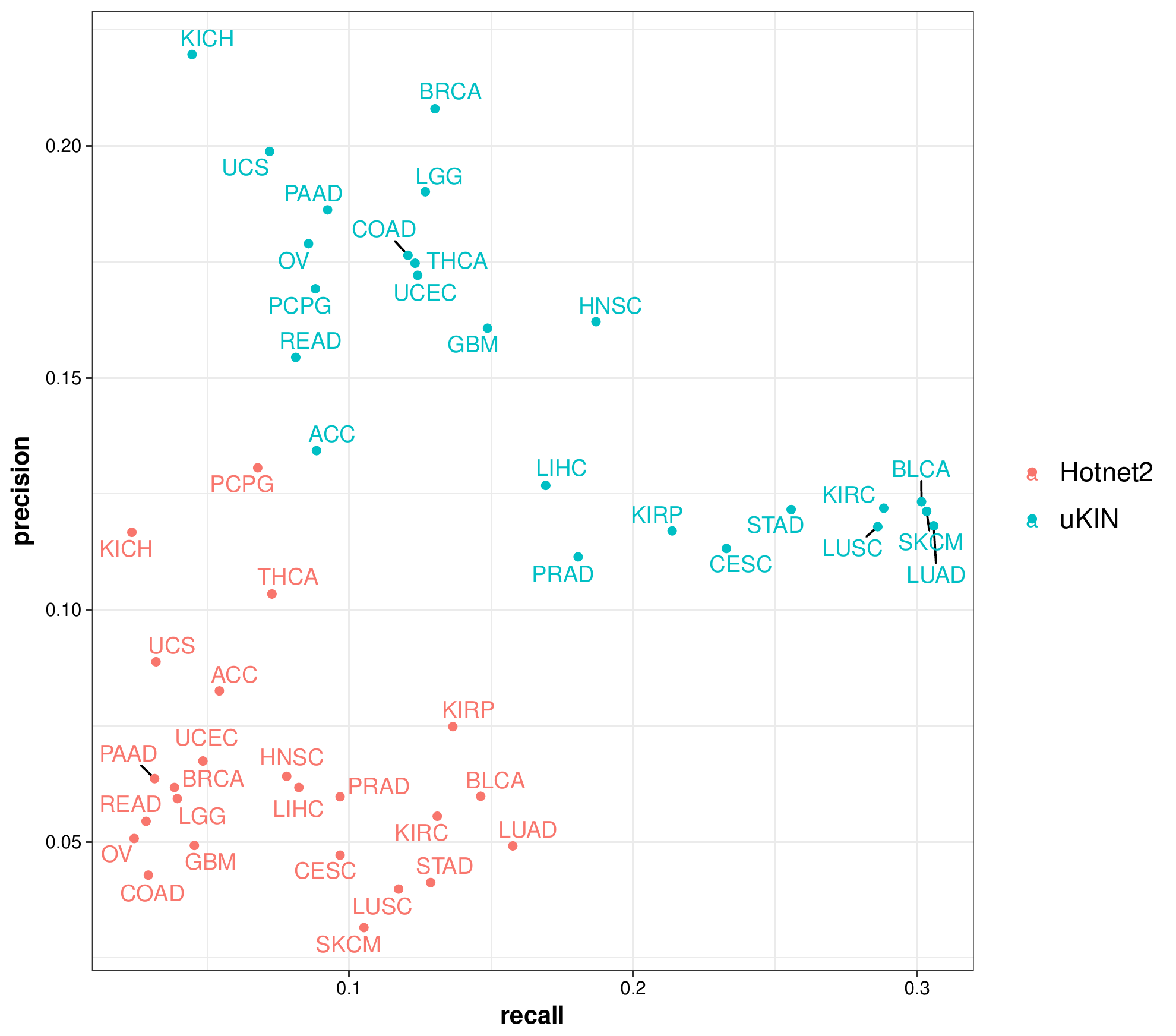}

\caption[Comparison between uKIN and Hotnet2]
	{{\bf Comparison between {\tt uKIN} and {\tt Hotnet2}.}  For
	each cancer type, we compute the precision and recall of the
	genes returned by {\tt uKIN} with $\alpha$=0.5 and {\tt Hotnet2}.  {\tt Hotnet2}
	is run with default parameters (100 permuted
	networks, and $\beta$ = 0.2 for the restart probability for the
	insulated heat diffusion process).  {\tt Hotnet2}
	outputs a set of genes predicted to be cancer-relevant, and these genes 
	are not ranked. Thus, for {\tt uKIN}, we consider the same number of top scoring
	genes as output by {\tt Hotnet2}.  {\tt
	uKIN} exhibits both higher precision and higher recall
	than {\tt Hotnet2} across all 24 cancer types.
	}  \label{fig:ukin_suppl:sfig_ukin_hotnet}
\end{figure}


\begin{figure}[t!]
  \includegraphics[width=1\textwidth,keepaspectratio]{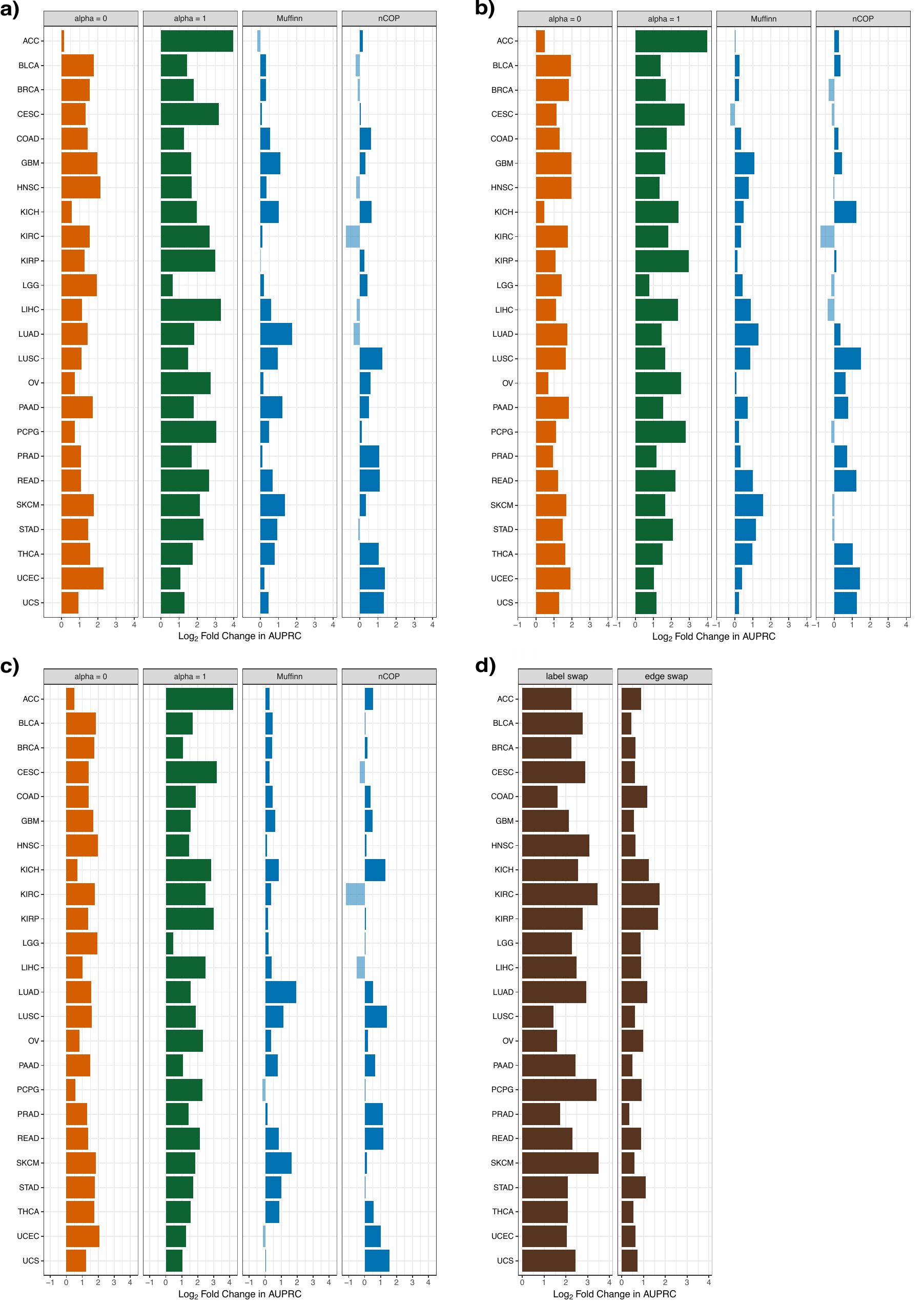}
\end{figure}

\clearpage
\newpage

\captionof{figure}[Robustness of uKIN]
	{{\bf Robustness of {\tt uKIN}.}  {\bf (a)} To make sure that
		 the results reported for {\tt uKIN} in Figures~2
		 and~3 are robust with respect to the set of labelled
		 cancer genes ${\cal H}$, instead of randomly sampling
		 400 genes from the Cancer Gene Census (CGC) list, we
		 form ${\cal H}$ using genes from other sources.
		 Specifically, we aggregate the cancer genes provided
		 by Hofree et al.\ in \cite{hofree2016challenges}
		 (which they obtained by querying the UniprotKB
		 database for the keyword-terms \lq
		 proto-oncogene,\rq~\lq oncogene\rq { }and \lq
		 tumoursuppressor\rq { }gene) and Vogelstein et
		 al.\ \cite{Vogelstein2013}, excluding any genes
		 present in the set of prior knowledge ${\cal K}$.
		 Log-fold AUPRCs are computed as described in the main
		 text. The results are consistent with those shown in
		 Figures~2 and~3 based on the CGC list and show the
		 superior performance of {\tt uKIN} as compared to the
		 other methods in recapitulating known cancer genes.
		 {\bf (b)} To make sure that the results reported for
		 {\tt uKIN} in Figures~2 and~3 are robust with respect
		 to number of genes used in evaluation, we compute
		 AUPRCs using the top 50 predicted genes. The results
		 are consistent with those shown in Figures~2 and~3
		 which use the top 100 predicted genes, and show the
		 superior performance of {\tt uKIN} as compared to the
		 baselines and other methods in recapitulating known
		 cancer genes.  The results are also consistent when
		 computing AUPRC's using 150 genes (data not shown).
		 {\bf (c)} To make sure that our method is robust with
		 respect to the specific network utilized, we repeat
		 our entire analysis procedure
		 for {\tt uKIN} with $\alpha = 0.5$
		 using the Biogrid network. The results are consistent
		 with those shown in Figures~2 and~3, based on the
		 HPRD network.  {\bf (d)} To make sure our method
		 utilizes network structure appropriately, we also
		 consider performance of {\tt uKIN} on the
		 real HPRD network as compared to
		 randomized HPRD networks. In the
		 left panel, we use a node label shuffling
		 randomization where the network structure is
		 maintained but gene names are swapped (thereby genes
		 can have very different numbers of interactions in
		 the randomizations).   In the right
		 panel,  we use a
		 classic {degree-preserving
		 randomization (edge swapping)}. 
 For each of the 24 cancers, we
		 compute the $log_2$ ratio of the area under the
		 precision recall curve using {\tt uKIN} with $\alpha = 0.5$
		 on the real network and on the randomized network and
		 show the average over 10 different randomizations.
		 Performance, as expected, is worse for both
		 randomizations across all cancers. We note that
		 significant cancer-relevant information is retained
		 in these randomized networks.  In particular, in both
		 types of network randomizations, we maintain the
		 relationships between genes and the samples that they
		 are found to be somatically mutated in.  Thus, some
		 highly mutated CGC genes may still be output by {\tt
		 uKIN} when running on randomized
		 networks.}  \label{fig:ukin_suppl:sfig2}

\clearpage
\newpage	


\begin{figure}[tb]
        \includegraphics[width=1\textwidth]{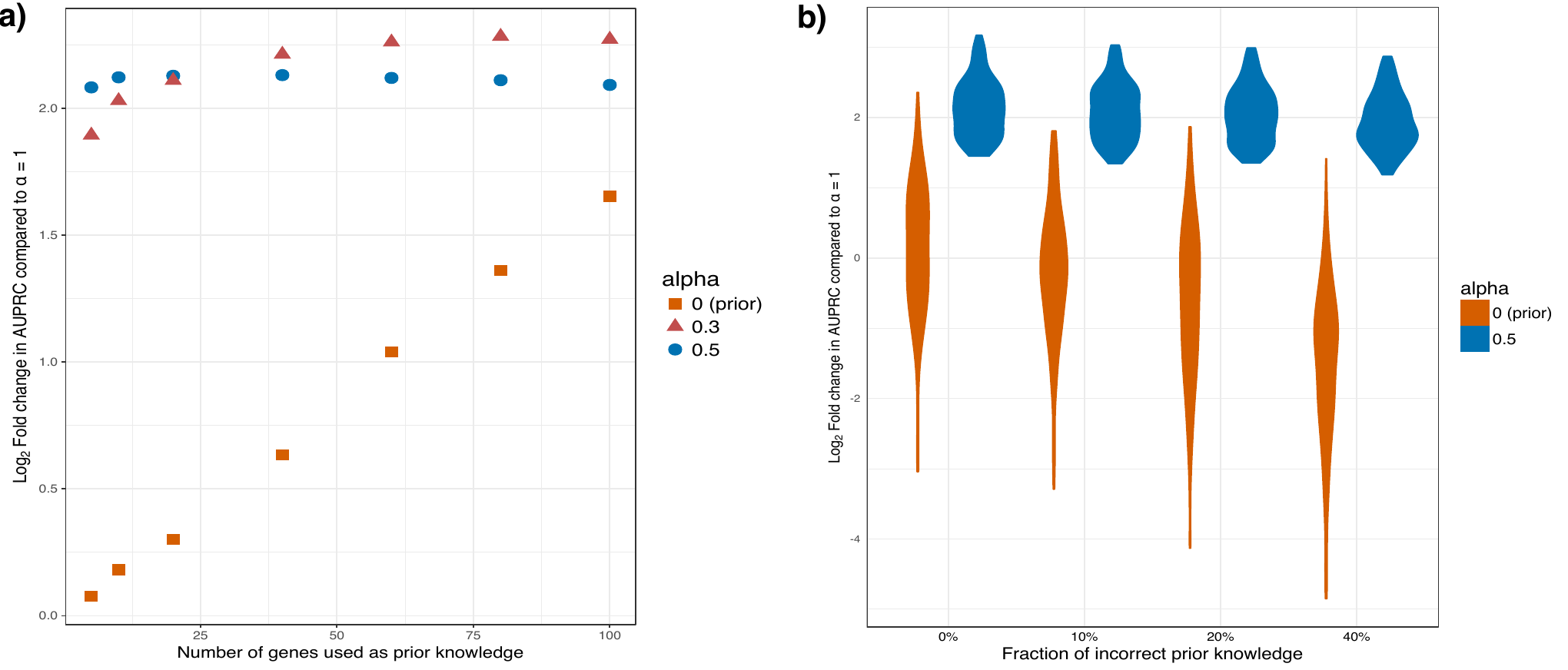}
        \caption[uKIN  benefits from larger and more accurate knowledge] {{\bf (a) {\tt
  uKIN} benefits from more knowledge.}  As we consider larger numbers
  of genes comprising the set of prior knowledge (${|\cal K}| = 5, 10
  ,20, 40, \dots, 100$), we examine the ability of {\tt uKIN} to
  uncover CGC genes in the same fixed set ${\cal H}$ when using
  $\alpha = 0.5$ (blue triangles), $\alpha = 0.3$ (pink circles) or
  $\alpha = 0$ (orange squares).  {\tt uKIN} is run on the HPRD network
  with the  kidney renal clear cell carcinoma (KIRC) dataset. We show the $\log_2$ ratio, averaged over 100
  runs, of the AUPRC of each version of {\tt uKIN} to the AUPRC for
  $\alpha=1$ which is constant across all possible ${\cal K}$ (and
  corresponds to the case where genes are ranked by mutational
  frequency).  For small ${\cal K}$, $\alpha=0$ performs poorly as is
  expected; as the prior knowledge available increases so does the
  performance.  For both $\alpha=0.3$ and $\alpha=0.5$, an increase in
  the size of ${\cal K}$ leads to an initial increase in the
  performance but eventually performance plateaus.  When limited prior
  knowledge is available ($|{\cal K}| < 20$), $\alpha=0.5$, which uses
  more of the new information, does better then $\alpha=0.3$, which
  relies more on using prior knowledge. When prior knowledge is
  abundant ($|{\cal K}| > 40$), {\tt uKIN} with $\alpha=0.3$
  outperforms $\alpha=0.5$.  As the number of genes comprising
the set of prior knowledge increases, spreading information just
from those genes ($\alpha=0$) improves in performance. This is consistent with the
observed clustering of CGC genes within biological
networks~\cite{Netbox2010}. However, even when propagating information from
100 known cancer genes, the performance is worse than
that when integrating it with new information (with either $\alpha=0.3$ or $\alpha=0.5$,
Figure~3a). {\bf (b) {\tt uKIN} is robust to small
  amounts of erroneous knowledge.}  We replace a fraction of the CGCs
  in the set of prior knowledge genes ${\cal K}$ with non-cancerous genes
  chosen uniformly at random from the set of non-CGC genes in the
  network.  We consider the performance for {\tt uKIN} with $\alpha = 0$ and
  $\alpha = 0.5$ when 0\%, 10\%, 20\% and 30\% of the prior knowledge
  genes are replaced with non-cancer genes.  100
  randomizations are performed at each level of incorrect
  knowledge. For each run, performance is measured as the $\log_2$
  ratio of the AUPRC of {\tt uKIN} (with either $\alpha = 0$ or $\alpha = 0.5$) to the AUPRC for the case where
  {\tt uKIN} is run with $\alpha=1$ (which is constant). {\tt uKIN} is run on the HPRD network
  with KIRC dataset with 20 CGC genes comprising the prior knowledge.
  Violin plots of this measure are shown are shown for $\alpha = 0$
  (orange) and $\alpha = 0.5$ (blue), jittered around the 0\%, 10\%,
  20\% and 30\% tick marks. At $\alpha = 0.5$, while performance steadily decreases, {\tt uKIN} remains
  robust to some incorrect knowledge {($\le 20\%$).}
  As expected, for $\alpha=0$, the
  decrease is more notable even when 10\% of the prior knowledge is
  incorrect because in that case {\tt uKIN} uses only prior knowledge.
}  
   \label{fig:ukin:fig5}
\end{figure}
\clearpage
\newpage	

	
\begin{figure}[hb]
  \includegraphics[width=0.7\textwidth,keepaspectratio]{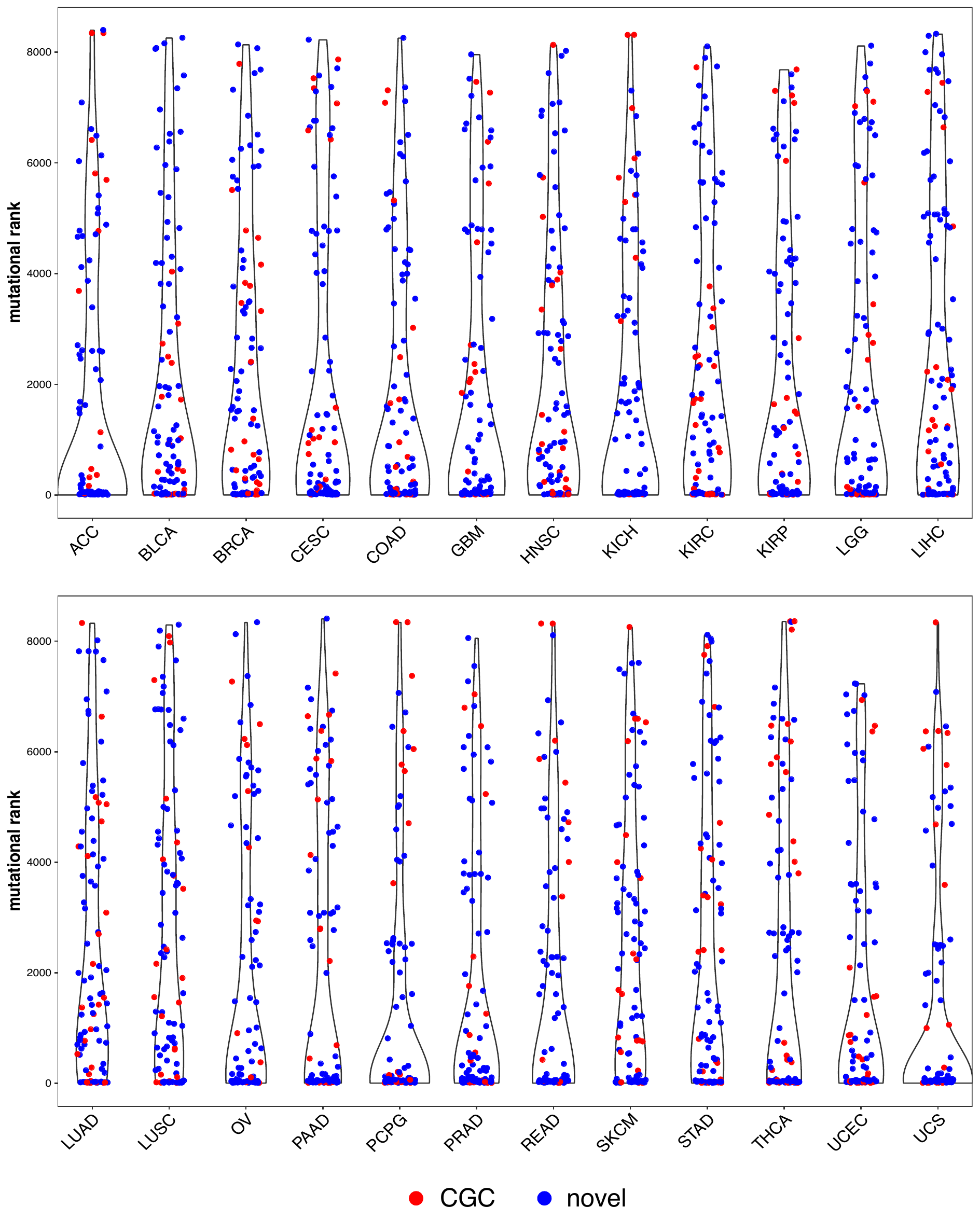}
\end{figure}
\captionof{figure}[uKIN identifies rarely mutated genes]
	{{\bf {\tt uKIN} identifies rarely mutated genes.}  To
illustrate {\tt uKIN}'s ability to predict genes as cancer-relevant 
cancer even if they are mutated across fewer numbers of individuals, we
consider mutation rates of {\tt uKIN}'s top scoring genes.  For each
cancer type, we run {\tt uKIN} 100 times with $\alpha=0.5$  and 20
genes as prior knowledge (see {\bf Methods}). For each
gene, its final score is obtained by averaging its scores (arising from the stationary distributions) across the runs; if a gene is in the set of prior knowledge genes ${\cal K}$ for a run, this run is not considered for its final score. For each of 
the $100$ genes with highest final scores, we consider the rank of its mutation rate
($y$-axis).  The mutation rate of a gene is
computed as the number of observed somatic missense and nonsense mutations across
tumors of that cancer type, divided by the number of amino acids in
the encoded protein. Then, for each cancer type, genes are ranked by
mutation rate where the gene with the highest mutation rate is given
the lowest rank. Known CGC genes are in red and novel predictions in
blue.  The top predictions consist of many heavily mutated genes
(i.e., those with low ranks), but {\tt uKIN} is also able to uncover
known cancer genes with very low mutational ranks (red dots towards
the top).  } \label{fig:ukin_suppl:sfig_violin_all_c}